\documentclass[a4paper]{article}
\usepackage{amssymb}
\usepackage{amsmath}
\usepackage{latexsym}
\usepackage{graphics}
\usepackage[dvips]{graphicx}
\def\setR{\mathbb{R}} 

\def\setZ{\mathbb{Z}} 
 
\def\proj{\mathbb{P}} 
 
\def\ie {i.e.} 

\DeclareMathAlphabet{\mcal}{OT1}{rsfs}{m}{sl}
\def\sgn{\textrm{sgn}}
\newcommand{\norm}[1]{\parallel\!#1\!\parallel}

\newcommand{\e}[1]{e^{#1}} 
\newcommand{\sss}[1]{\scriptscriptstyle #1} 

\begin{document}
%\preprint{APS/123-QED}

\title{Conformally related massless fields in dS, AdS and Minkowski spaces}

%\begin{center}
\author{
E. Huguet$^{1,2}$, J. Queva$^{1,2}$, J. Renaud$^{1,3}$
}

%\email{huguet@apc.univ-paris7.fr, jacques.renaud@univ-mlv.fr}

\maketitle
{\small 
\noindent $1$ - APC, 11 pl. M. Berthelot, F-75231 Paris Cedex 05, France.  \\
$2$ - Universit\'e Paris 7 Denis Diderot, boite 7020, F-75251 Paris Cedex 05, France.
\\
$3$ - Universit\'e de Marne la Vall\'ee, 5 Bld Descartes, Champs sur marne, 77454 Marne-la-Vall\'ee Cedex 2, France.}\\
{\footnotesize e-mail: eric.huguet@apc.univ-paris7.fr, jacques.renaud@univ-mlv.fr }
\begin{abstract}
In this paper we write down the equation for a scalar conformally coupled field simultaneously 
for de Sitter (dS), anti-de Sitter (AdS) and Minkowski spacetime in $d$ dimensions. The curvature dependence appears in a very simple way through a conformal factor. As a consequence the process of curvature free limit, including  wave functions limit and two-points functions, turns to be a straightforward issue. We determine a set of modes, that we call de~Sitter plane waves, which become ordinary plane waves when the curvature vanishes.
\end{abstract}

%\pacs{04.62.+v, 98.80.Jk}% PACS, the Physics and Astronomy

\section{Introduction}
The crux of this paper is the following. We write down the de~Sitter (dS), anti-de~Sitter (AdS) and Minkowski spacetimes on the same underlying set where we identify a system of coordinates which corresponds to ordinary coordinates on Minkowski spacetime, and we obtain dS conformally invariant objects such as plane waves and two-point functions written in term of Minkowski coordinates with a convenient dependence on the curvature.

The de~Sitter, Anti-de~Sitter  and Minkowski spaces are known to be intimately related: they are 
maximally symmetric spaces the two former being the only homogeneous and isotropic deformations
of the latter. Also their respective isometry group SO$_0(1,d )$ (dS),  
SO$_0(2,d - 1)$ (AdS) and $d$-dimensional Poincar\'e group (Minkowski space) are subgroup of 
the Conformal group SO$_0(2,d)$.

Despite the above nice characteristics linking the de~Sitter, Anti-de~Sitter  and Minkowski 
spaces, things are not simple when dealing with field theory \cite{birreldavis, fulling, wald, bk}. 
Amongst other things the relation between fields on dS or AdS spaces and fields on Minkowski space, 
when such a relation exists, may require a not so obvious procedure. This, in particular, is the case for the conformally coupled massless scalar field (CCMSF) in dS and Minkowski spaces. In that case one generally resorts to group theory \cite{barutraczka,fronsdal,onofri}: the unitary irreducible representations corresponding to the CCMSF in dS and Minkowski spaces are extended to the conformal group where they can be identified. The drawback of this method is that the transformation of the field is not apparent. On another hand one can perform the
zero curvature limit on the field but this is not a straightforward procedure. Moreover, there is no 
easy way to solve the inverse problem: inserting curvature into Minkowskian objects.

In the present work, we handle the CCMSF simultaneously on AdS, dS and Minkowski spaces (see \cite{zg} for a similar approach) and, as a byproduct,
give a new solution to the above problem of correspondence between fields on dS and Minkowski spaces. We 
present a set of modes, that we call de~Sitter plane waves. These modes verify the de~Sitter Klein-Gordon 
equation and depend explicitely on the curvature, they become the ordinary plane waves when curvature is set 
to zero.

In fact since AdS, dS and Minkowski spacetimes are conformally related, we write down these spaces on the same underlying set. This allows to deform Minkowskian objects into de Sitterian objects, precisely, we can describe every dS object in term of usual Minkowskian coordinates. This procedure applies, 
in particular, to two-points functions. Note that, starting from a group theoretical point of view, Onofri \cite{onofri} found a set of modes on the projective cone which seems to us difficult to interpret as spacetime functions. Our modes differ from those of Onofri through a phase factor which allows this interpretation.

Our paper is organized as follows. The AdS, dS and Minkowski spaces are obtained 
as intersections of the $(d + 1)$-dimensional cone by a moving plane, the position of the plane relatively to the cone determining the nature of the space. We thus can go continuously from one space to another by changing the position of the plane. This is done in  Sec.\ \ref{XH}.  We show in Sec.\ \ref{Cprime}  that the above intersection can be identified with a subset of the $(d + 1)$-dimensional cone up to dilations $\mathcal{C}'$. The solution of the CCMSF equation on this subset is then obtained and the solutions in AdS, dS and Minkowski spaces follows (Sec.\ \ref{solsurXH}). In Sec.\ \ref{dsmink} we study the link between the solutions on dS and Minkowski spaces. 
In particular we show that their respective Hilbert spaces are related by an unitary map. We then identify de Sitter plane waves and
calculate the two-points function, which is found to have the Hadamard behavior.  The dimension $d = 2$ requires a special attention and is examined 
in Sec.\ \ref{d=2}. 

\subsubsection*{Conventions}
Here are the conventions about indices:
\begin{eqnarray*}
\alpha, \beta, \gamma, \delta, \ldots &=&d+1, 0, \ldots, d\\
\mu,\nu,\rho,\sigma,\ldots &=&  0, \ldots, d-1\\
i, j, k, l, \ldots &=& 1, \ldots, d-1.
\end{eqnarray*} 
The coefficients of the metric $diag(1,1,-1,\ldots,-1)$ of $\setR^{d+2}$ are denoted $\eta_{\alpha \beta}$:
\begin{equation}
\eta_{d+1\ d+1}=\eta_{00}=1=-\eta_{ii}=-\eta_{dd}.
\end{equation}

\section{The manifold $X_{\xi}$}\label{XH}
In this section we show how to obtain de~Sitter, Minkowski and anti-de~Sitter spaces by intersecting the null cone with a moving plane.

We consider  $\setR^{d+2}$ provided with the metric $ds^2=\eta_{\alpha\beta}y^\alpha y^\beta$. Let 
\begin{equation}\label{C}
\mathcal{C} = \left\{y \in \setR^{d+2} :  (y^{d+1})^2 + (y^0)^2 - \boldsymbol{y}^2 - (y^d)^2 = 0\right\}, 
\end{equation}
be the null cone of $\setR^{d+2}$. 
Now, let $\xi$
be a real parameter and consider 
the plane
\begin{equation}\label{PH}
P_{\xi} = \left\{y \in \setR^{d+2} : (1 + \xi) y^{d+1} + (1 - \xi) y^{d} = 2 \right\}. 
\end{equation}
We set
\begin{equation}\label{defdsH}
X_{\xi} = \mathcal{C} \cap P_{\xi}.
\end{equation}

Now, let us show that $X_{\xi}$ reduces to a dS (respectively AdS, Minkowski) space for $\xi > 0 $ (respectively $\xi < 0$, $\xi = 0$). First note that the manifold $X_{\xi}$ has a constant Ricci scalar given by $R_{\xi}=-d(d-1)\xi$, with respect to the natural metric inherited from the ambient space. This can be checked by  a direct calculation.

Secondly, let us show that $X_{\xi}$ is in fact invariant under a de Sitter (respectively AdS, Poincar\'e) subgroup of SO$(2,d)$ parameterized by $\xi$. 
The generators of the algebra so$(2,d)$ are 
$X_{\alpha \beta} = y_\alpha \partial_\beta -  y_\beta \partial_\alpha$, they satisfy
\begin{equation}\label{so2d}
\left[X_{\alpha \beta}, X_{\gamma \delta}\right] =  \eta_{\beta \gamma} X_{\alpha \delta} +\eta_{\alpha \delta} X_{\beta \gamma} 
- \eta_{\alpha \gamma} X_{\beta \delta} - \eta_{\beta \delta} X_{\alpha \gamma}. 
\end{equation}  
Obviously the $(d-1)(d-2)/2$ generators $X_{ij}$ and the $d-1$ generators $X_{0i}$ of so$(2,d)$ leave $X_{\xi}$ invariant.  The $d$ more generators which also leaves $X_{\xi}$ invariant  
are found to be
\begin{equation}\label{yh}
Y^{\xi}_\mu := \frac{1}{2} (1 - \xi) X_{d+1 \mu} + \frac{1}{2} (1 + \xi) X_{\mu d}.
\end{equation}
A straightforward calculation leads to the following commutations relations:
\begin{eqnarray}\label{dsh-algebra}
\left[Y^{\xi}_\mu, Y^{\xi}_\nu\right] &=&  \xi  X_{\mu \nu}, \\
\left[ Y^{\xi}_\rho, X_{\mu \nu} \right] &=& \eta_{\mu \rho} Y^{\xi}_\nu - \eta_{\nu \rho} Y^{\xi}_\mu. 
\end{eqnarray}
The remaining commutation relations between the other generators $X_{\mu \nu}$ leaving $X_{\xi}$ invariant satisfy the Lorentz so$(1,d-1)$ algebra:
\begin{equation}\label{so1d}
\left[X_{\mu \nu}, X_{\rho \sigma}\right] = \eta_{\nu \rho} X_{\mu \sigma} +\eta_{\mu \sigma} X_{\nu \rho} 
- \eta_{\mu \rho} X_{\nu \sigma} - \eta_{\nu \sigma} X_{\mu \rho}. 
\end{equation}  
For $\xi = H^2 > 0$ (respectively $\xi = 0$, $\xi = - H^2 < 0$) the above relations are the familiar algebra of the de Sitter 
(respectively Poincar\'e, AdS) group. Now, varying continuously the parameter $\xi$ between $-\infty$ and $+\infty$ 
leads to a continuous deformation of the above algebra.

One can finally, for the sake of completeness, give an equation for $X_{\xi}$ which leads to that of 
a dS, AdS or Minkowski space for the corresponding value of $\xi$. To this end we  change the variables from $(y^0,\boldsymbol{y},y^{d},y^{d+1})$ to $(y^0,\boldsymbol{y},v,w)$ where 
\begin{eqnarray}
v &=& \frac{1}{2 }(1 + \xi)y^{d+1}  + \frac{1}{2 } (1 - \xi) y^d,\\
w &=& \frac{1}{2 }(1 - \xi)y^{d+1}  + \frac{1}{2 } (1 + \xi) y^d,\label{w}
\end{eqnarray}
for $\xi \ne 0$.
Then  the equation for $X_{\xi}$ reads
\begin{equation}\label{C1}
(y^0)^2 - \boldsymbol{y}^2 -  \frac{w^2}{\xi} = - \frac{1}{\xi},
\end{equation}
which reduce to the usual equation for a dS (respectively AdS) space if one sets $\xi = H^2 > 0$ 
(respectively $\xi = - H^2  < 0$) and ${\displaystyle u = \frac{w}{H}}$.
 To handle the case $\xi = 0$ note that the above equation reads 
$\xi\left((y^0)^2 - \boldsymbol{y}^2\right) -  w^2 + 1 = 0$ which for $\xi = 0$, shows that $y^\mu$ are 
free to vary in $\setR$ and that consequently $y$ belongs to a $d$-dimensional Minkowski space.

\section{The cone up to dilations}\label{Cprime}
We now compute the metric of $X_{\xi}$ in a convenient 
coordinate system on the cone $\mathcal{C}$:
\begin{equation}\label{coordC}
\left \{
 \begin{array}{lcl}
  y^{d+1} &=& r_c \cos \beta \\
  y^0 &=& r_c \sin \beta \\
  y^i &=& r_c \sin \alpha\; \omega^{i}(\theta_1,\ldots,\theta_{d-2})\\
    y^d &=& r_c \cos \alpha,
 \end{array}
\right .
\end{equation}
in which $r_c\in[0,+\infty[,\, \beta\in[-\pi,\pi[,\, \alpha\in[0,\pi]$ and $\omega^{i}(\theta_1,\ldots,\theta_{d-2})$ belongs to $S^{d-2}\subset\setR^{d-1}$.
The definition (\ref{PH}) of the plane $P_\xi$ leads to the constraint
\begin{equation}\label{rconstraint}
r_c \bigl\vert_{\sss P_{\xi}} = \Omega_{\xi}(\alpha,\beta),
\end{equation}
where we defined
\begin{equation}\label{romega}
\Omega_{\xi}(\alpha,\beta) := \frac{2}{(1 - \xi)\cos\alpha + (1 + \xi)\cos\beta}.
\end{equation}
Note that since $r_c$ is always positive the same is true for $\Omega_{\xi}(\alpha,\beta)$.
Taking into account the above constraint into the coordinate system (\ref{coordC}) we deduce the line element on $X_{\xi}$:
\begin{equation}\label{ds2dsh}
ds_{\xi}^2 = \Omega_{\xi}^2(\alpha, \beta) ds^2, 
\end{equation}
where $ds^2 = d\beta^2 - d\alpha^2 - \sin^2\!\!\alpha\, d\omega_{\sss d-2}^2$, $d\omega_{\sss d-2}$ being
 the line element of the $(d-2)$-sphere.  
 
At this point different values of $\xi$ correspond to different $X_{\xi}$ manifolds which are all different submanifolds of the cone $\mathcal{C}$. Nevertheless, the formula (\ref{ds2dsh}) suggests that these spaces can be  now realized on the same manifold but with different metrics, depending on $\xi$ through the conformal factor.
To this end let us introduce the cone up to dilations 
\begin{equation} \mathcal{C}'=\mathcal{C}/\sim,\end{equation}
where the relation $\sim$ is defined through $u\sim v$ if and only if 
there exists $\lambda>0$ such that $u=\lambda v$.  
The cone $\mathcal{C}$ is left invariant under the natural action of the group SO$(2,d)$ in $\setR^{d+2}$.
This action extends to $\mathcal{C}'$ which is also left invariant. 
Thus, $\mathcal{C}'$ can be considered as the submanifold $r_c = 1$ of $\mathcal{C}$ and we will adopt this point of view in the sequel, therefore convenient coordinates on 
$\mathcal{C}'$ are
\begin{equation}\label{coordCp}
\left \{
 \begin{array}{lcl}
  y^{d+1} &=&  \cos \beta \\
  y^0 &=&  \sin \beta \\
  y^i &=&  \sin \alpha\; \omega^{i}(\theta_1,\ldots,\theta_{d-2})\\
  y^d &=&  \cos \alpha.
 \end{array}
\right .
\end{equation}
In view of these coordinates, one has immediately $\mathcal{C}'\simeq S^1\times S^{d-1}$. Note that $\mathcal{C}'$ is {\em not} the projective cone in which $y$ and $-y$ should be identified.

We can now consider $X_{\xi}$ as a subset of $\mathcal{C}'$ through the identification $\sim$. This subset turns out to be defined through
\begin{equation}
X_{\xi}= \left\{y \in \mathcal{C}' : (1 + \xi) y^{d+1} + (1 - \xi)y^d > 0\right\}. 
\end{equation}
The metric on $X_{\xi}$ is still given by (\ref{ds2dsh})
where  $ds^2$ is precisely the natural line element of $\mathcal{C}'$,~\ie~induced from $\setR^{d +2}$. 
The metric on X$_{\xi}$ and on $\mathcal{C}'$ are thus
conformally related. Note that subsequently the same is true for different values of $\xi$, that is
$X_{\xi_1}$ and $X_{\xi_2}$ with $\xi_1\neq \xi_2$ are also conformally related.

\begin{figure}[ht]
\begin{center}
\includegraphics[width = 8cm]{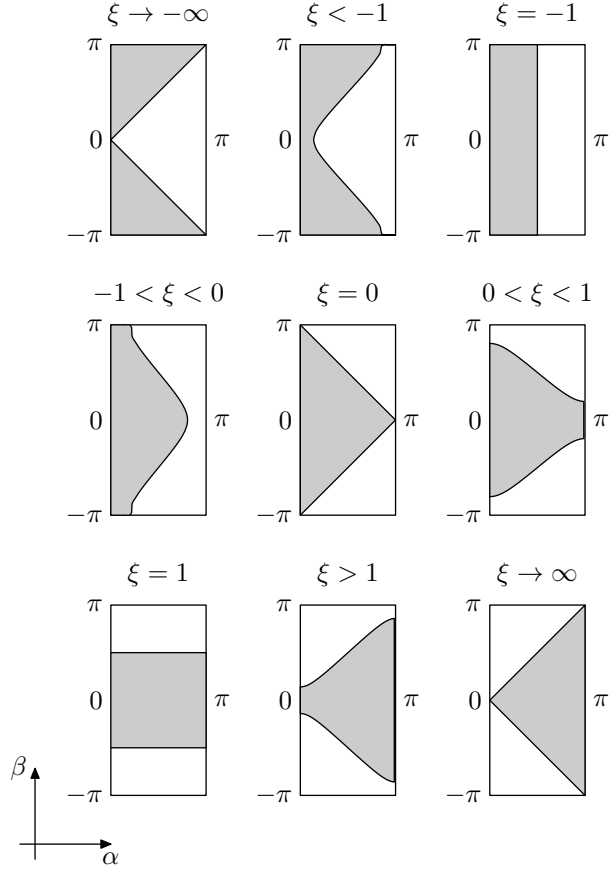}%mps
\caption{The set $X_\xi$, shaded region, as a subset of the cone up to dilations $\mathcal{C}'$. As usual the angular coordinates do not appear so that each point corresponds to a $S^{d - 2}$ sphere. 
The four first diagrams correspond to AdS space, the fifth is the Minkowski space and the four last correspond 
to dS space.}\label{DiagPenrose5}
\end{center}
\end{figure}

Finally the manifolds $X_\xi$ are displayed in Fig.\ \ref{DiagPenrose5}.  

\section{The conformally coupled massless scalar field on $X_{\xi}$}\label{solsurXH}

We can now deal with de Sitter, anti-de Sitter and Minkowski spaces at the same time: we just  have to consider $\mathcal{C}'$ and use the conformal correspondence described above. As an application, let us obtain the solutions of the field equation
for a  conformally coupled massless scalar field (CCMSF) on $X_{\xi}$.
The CCMSF equations for respectively $\mathcal{C}'$ and $X_{\xi}$ reads 
\begin{eqnarray}
(\square + \frac{1}{4} (d-2)^2) \phi &=& 0, \label{eqpc}\\
(\square_{\xi} +  \frac{1}{4} d (d-2) \xi) \phi^{\xi} &=& 0, \label{eqdsh}
\end{eqnarray}
with
\begin{eqnarray}
\square &=&  \partial^2_{\beta\beta} - \Delta_{d-1},\label{dalC'}\\
\square_{\xi}&=&
\frac{1}{\Omega_{\xi}^2} \square + \frac{d-2}{2}\frac{1}{\Omega_{\xi}} \times \nonumber\\
&\times&\left(
(1+\xi)\sin\beta \partial_\beta
-(1-\xi)\sin\alpha \partial_\alpha\right),\label{dalH}
\end{eqnarray}
where $\Delta_{d-1}$ is the Laplacian on $S^{d-1}$. It is worth noting that the operator $\square_{\xi}$ (\ref{dalH}) will not be used in calculations, it is given for completeness. Thanks to the conformal relation between $X_{\xi}$ and $\mathcal{C}'$, the 
solutions $\phi$ and $\phi^{\xi}$ of CCMSF equations on respectively $\mathcal{C}'$ and $X_{\xi}$  are related through (see for instance \cite{birreldavis,bk})
\begin{equation}\label{phi-phih} 
\phi^{\xi} = \Omega_{\xi}^{-(\frac{d - 2}{2})} \phi.
\end{equation}

The solution of (\ref{eqpc}) can easily 
be found by the method of the separation of variables, it reads
\begin{equation}\label{sol-eqpc4}
\phi_{{\sss L} {\sss M}}(\beta,\alpha,{\boldsymbol{\theta}}) =C_{Ld} \e{-i(L + \frac{d-2}{2}) \beta} \mathcal{Y}_{{\sss L} {\sss M}}(\alpha,{\boldsymbol{\theta}}),
\end{equation}
where ${\boldsymbol{\theta}}= (\theta_1,\ldots,\theta_{d-2})$ and $\mathcal{Y}_{{\sss L}  {\sss M}}(\alpha,{\boldsymbol{\theta}})$ denotes usual hyperspherical harmonics \cite{Hyp.Geom.85}, which are an orthonormal basis of L$^2(S^{d-1})$ fulfilling
$$\Delta_{d-1}\mathcal{Y}_{{\sss L}  {\sss M}}=-L(L+d-2)\mathcal{Y}_{{\sss L}  {\sss M}},$$
and $C_{{\sss L}d}$ is a normalization constant to be discussed below.
The solution of (\ref{eqdsh}) are thus
\begin{equation}\label{sol-eqdsh4}
\phi^{\xi}_{{\sss L}{\sss M}}(\beta,\alpha,{\boldsymbol{\theta}}) =  \Omega_{\xi}^{-(\frac{d-2}{2})}
C_{{\sss L} d} \e{-i(L + \frac{d-2}{2}) \beta} \mathcal{Y}_{{\sss L} {\sss M}}(\alpha,{\boldsymbol{\theta}}).
\end{equation}
Note that the dependence on $\xi$ allows to set $\xi = 0$ without any complication. As shown in Sec.\ V  below, this set of modes yields, for $\xi=0$, the usual Hilbert space of positive frequency solutions of the Minkowskian Klein-Gordon equation.

One can compare our set of solutions with those of Onofri \cite{onofri, jpg}.
The $\mathcal{Y}_{{\sss L}  {\sss M}}(\alpha,{\boldsymbol{\theta}})$ can be seen as homogeneous 
polynomials of degree $L$ on $\setR^d$ and (\ref{sol-eqpc4}) can be recast as follows:
\begin{equation}
\phi_{{\sss L}{\sss M}}=
 \e{-i( \frac{d-2}{2}) \beta} \mathcal{Y}_{{\sss L} {\sss M}}(e^{-i\beta} z),
\end{equation}
where $z \in\setR^d$. The phase term before $\mathcal Y$ is not present in Onofri's paper, and this term forbids the identification of $(\beta,z)$ with $(\beta+\pi,-z)$, except for $d\equiv 2 \,[\mbox{modulo } 4]$. In other words we must work on $\mathcal{C}'\simeq S^1\times S^{d-1}$ and not on the projective cone $\proj \mathcal{C}\simeq S^1\times S^{d-1}/\setZ_{\sss 2}$, except for $d\equiv 2\, [\mbox{modulo } 4]$.

\section{Relation between fields on de Sitter and \\Minkowski spaces.}\label{dsmink}
In this section we restrict our study to the globally hyperbolic spaces, namely de~Sitter and Minkowski 
spaces: $\xi = H^2$, $H$ being a non-negative real 
parameter, and we assume that $d>2$. Accordingly, we slightly modify our notations in this section: the quantities previously indexed 
by $\xi$ become indexed by $H$, thus for instance we denote by $X_{\sss H}$ the manifold 
described in Sec.\ \ref{XH}. 

We study the correspondence $\phi\to \Omega_{\sss H}^{-(\frac{d-2}{2})} \phi$ from the Hilbert space point of view, define what we call the de~Sitter plane waves and  write down the two-points function.

\subsection{Scalar Products}
Let $\check{\mathcal{C}}'$ be the space $\mathcal{C}'$ with a cut at $\beta=\pi$: 
$$\check{\mathcal{C}}'=]-\pi,\pi[\times S^{d-1}.$$
The spaces $\check{\mathcal{C}}'$ and $X_{\sss H}$ are both globally hyperbolic, the space-like hypersurface $\beta=0$ being a common Cauchy hypersurface. Their respective Klein-Gordon scalar products read
\begin{equation}\langle\phi_1|\phi_2\rangle_{\check{\mathcal{C}}'}= i \int_{\beta=0}\phi_1^*
\stackrel{\leftrightarrow}{\partial_\beta}\phi_2\,dv_{\sss d-1},
\end{equation}
on $\check{\mathcal{C}}'$ and
\begin{equation}\langle\phi_1^{\sss H}|\phi_2^{\sss H}\rangle_{\sss H}= i\int_{\beta=0}(\phi_1^{\sss H})^*
\stackrel{\leftrightarrow}{\partial_\beta}\phi_2^{\sss H}\;\Omega_{\sss H}^{d-2}dv_{\sss d-1},
\end{equation}
on $X_{\sss H}$, where  $dv_{\sss d-1}$ is the volume element of $S^{d-1}$.
Let ${\cal H}'$ (respectively $\mathcal{H}_{\sss H}$) be the Hilbert space of the solutions $\phi$  
(respectively $\phi^{\sss H}$) of 
(\ref{eqpc}) (respectively 
(\ref{eqdsh})) on $\check{\mathcal{C}}'$ (respectively  $X_{\sss H}$) which verify 
$\langle\phi|\phi\rangle <\infty$ (respectively $\langle\phi^{\sss H}|\phi^{\sss H}\rangle <\infty$) and let $\widehat{\Omega}_{\sss H}$ be the operator
\begin{equation}\label{omegamap}
\begin{array}{lrcl}
\widehat{\Omega}_{\sss H}: &{\cal H}'&\to& \mathcal{H}_{\sss H}\\
&\phi&\mapsto& \widehat{\Omega}_{\sss H}(\phi) := \Omega_{\sss H}^{-(\frac{d-2}{2})}\phi.
\end{array}
\end{equation}
A straightforward calculation shows that $\widehat{\Omega}_{\sss H}$ is unitary with respect to the scalar products in $\check{\mathcal{C}}'$ and $X_{\sss H}$,~\ie:
\begin{equation} \langle\phi_1|\phi_2\rangle_{\check{\mathcal{C}}'}=\langle\widehat{\Omega}_{\sss H}(\phi_1)|
\widehat{\Omega}_{\sss H}(\phi_2)\rangle_{H}.
\end{equation}
Now the definition of a scalar product on $\mathcal{H}'$ determines the normalization 
constant in (\ref{sol-eqdsh4}) which is found to be
\begin{equation}\label{eq.Norm}
C_{{\sss L}d}= \left(2L + d-2\right)^{-\frac{1}{2}}\, .
\end{equation}

It is worth noting that in (\ref{omegamap}) $\phi$ is defined on $\check{\mathcal{C}}'$ whereas $\widehat{\Omega}_{\sss H}(\phi)$ is only defined on $X_{\sss H}$. Thus,
it may be some subsets of $\check{\mathcal{C}}'$ where a function which belongs to $\mathcal{H}'$ 
is not defined. Nevertheless, such a situation is not problematic since all the functions 
are defined on the same Cauchy hypersurface ($\beta = 0$) and  $\widehat{\Omega}_{\sss H}$ always maps 
a solution of (\ref{eqpc}) to a solution of (\ref{eqdsh}) and thus is well defined and invertible.  

The group acting on de Sitter and Minkowski spacetimes are different, as a consequence, there is no meaning to speak about covariance of $\widehat{\Omega}_{\sss H}$ with respect to these groups. Nevertheless, this operator is obviously covariant with respect to the subgroup generated by $X_{ij}$ and $X_{0i}$. This subgroup can be interpreted, from the Minkowskian point of view, as a Lorentz group $X_{0i}$ being the boosts and $X_{ij}$ the rotations. It is also the stabilizer of the point $\alpha=0,\beta=0$ at the vicinity of which the de Sitter quantities can be understood from a flat space point of view  \cite{ghr1}. For instance, $X_{0i}$ can be interpreted as a boost near this point, but not in the whole de~Sitter spacetime.

\subsection{Minkowskian coordinates and dS plane waves}

For $H = 0$, $X_{\sss H}$ becomes the $d$-dimensional Minkowski spacetime and one recover 
the usual Minkowski coordinates $\{x^\mu\}$ by setting
\begin{eqnarray}
 x^0 &=& \Omega_0 \sin\beta,\label{coordMink0}\\
 x^i &=& \Omega_0 \sin\alpha~\omega^i,\label{coordMinki}
\end{eqnarray}
where $\Omega_0 := \Omega_{{\sss H}=0}$. The global time coordinate $t$ and the radius $r$ of 
the $(d - 2)$-sphere are thus: $t = x^0$ and 
$r = \norm{\boldsymbol{x}} = \Omega_0 \sin\alpha$. These coordinates $\{x^\mu\}$ can also be considered as coordinates on ${\mathcal{C}}'$ and also on $X_{\sss H}$, although they do not cover the whole space. We call them $M$-coordinates in the sequel.

The usual Minkowskian planes waves $\e{- i k x}$, where $k$ 
is the $d$-momentum vector satisfying $\boldsymbol{k}^2 = (k^0)^2$ and 
$k x =k_\mu x^\mu$, can be transported
on $\mathcal{C}'$  giving new sets of solutions of (\ref{eqpc}):
\begin{equation}\label{expC'}
f'_k(\alpha,\beta,\boldsymbol{\theta}) = \Omega_0(\alpha,\beta)^{(\frac{d - 2}{2})} \e{- i k x},
\end{equation}
these functions give in turn a set of solutions of (\ref{eqdsh}):
\begin{equation}\label{expdsH}
f_k^{{\sss H}}(\alpha,\beta,\boldsymbol{\theta}) = 
\left(
\frac{\Omega_0(\alpha,\beta)}{\Omega_{\sss H}(\alpha,\beta)}
\right)^{(\frac{d - 2}{2})}\e{- i k x}.
\end{equation}

In the above expressions the coordinates $\{x^\mu\}$ are given by (\ref{coordMink0}-\ref{coordMinki}) which can be inverted to give
\begin{equation}\label{expC'M}
f'_k(x) =  \left(\frac{1}{4} \sqrt{(4 + N_+^2)
(4 + N_-^2)}\right)^{(\frac{d - 2}{2})} \e{- i k x},
\end{equation}
where
$
N_+ := (x^0 + \norm{\boldsymbol{x}}),\;
N_- := (x^0 - \norm{\boldsymbol{x}})
$
and
\begin{equation}\label{expdsHM}
f_k^{{\sss H}}(x) = 
 \left(1 - \frac{H^2}{4^{~}}x^2\right)^{(\frac{d - 2}{2})} \e{- i k x},
\end{equation} 
which can be interpreted as the de Sitter plane waves (solutions of (\ref{eqdsh})) written in $M$-coordinates. We do insist on the fact that, since Minkowski and de~Sitter spacetimes are written on the same underlying set there is no problem in writing down de Sitter objects in Minkowskian coordinates.

\subsection{Two-points function}
Let $D_{\sss{H}}^{+}$ and $D_{\mathcal{C}'}^{+}$ be the Wightman two-points functions on $X_{\sss H}$ and ${\mathcal C}'$. They read
\begin{eqnarray}\label{eq.Wight1}
 D_{\sss{H}}^{+}(y, y')
       &:=& \sum_{\sss{LM}}\quad
                \phi^{\sss{H}}_{\sss{LM}}(y)\,\phi^{\!\ast\sss{H}}_{\sss{LM}}(y') 
                \nonumber\\
        &=& \Omega_{\sss{H}}^{-\left(\frac{d-2}{2}\right)}(y)     
                D_{\mathcal{C}'}^{+}(y,y')
           \Omega_{\sss{H}}^{-\left(\frac{d-2}{2}\right)}(y')   
\end{eqnarray}
where the functions $\phi^{\sss{H}}_{\sss{LM}}$ are those defined in \eqref{sol-eqdsh4} and
normalized by \eqref{eq.Norm}. Let us remind the formula \cite{Hyp.Geom.85}
\begin{equation}
\sum_{\sss{M}} \frac{
        \mathcal{Y}_{\sss{LM}}^{}(\alpha,\,\boldsymbol{\theta})
        \mathcal{Y}_{\sss{LM}}^{\ast}(\alpha',\,\boldsymbol{\theta'})}{2L + d - 2}
 = \frac{\Gamma\left(\frac{d-2}{2}\right)}{4\pi^{\frac{d}{2}}}\,
        C^{\frac{d-2}{2}}_{\sss{L}}(\cos\omega) \,,
\end{equation}
where $\omega$ is the angle between the vectors in
$S^{d-2}\subset\setR^{d-1}$ parameterized by $(\alpha,\,
\boldsymbol{\theta})$ and $(\alpha',\, \boldsymbol{\theta'})$, and $C$ a usual
Gegenbauer polynomial.
Inserting the above  equation in $D^{+}_{\mathcal{C}'}$  and adding a $\varepsilon$ term in order to make the series convergent, leads to the following expression:
\begin{multline}
 D^{+}_{\mathcal{C}'}(y, y')
        = \frac{\Gamma\left(\frac{d-2}{2}\right)}{4\pi^{\frac{d}{2}}}\,
                e^{-i \frac{d-2}{2} (\beta - \beta')}\,\times\\
                \times\sum_{\sss{L}=0}^{\infty}  e^{-i L \big[(\beta - \beta')
-i\varepsilon\big]}
                C^{\frac{d-2}{2}}_{\sss{L}}(\cos\omega)
\end{multline}
where the series in the rhs is  the boundary value of the generating function  for Gegenbauer
polynomials. The function $D^{+}_{\mathcal{C}'}(y, y')$ is then straightforwardly
obtained as
\begin{equation}\label{dcprime}
 D^{+}_{\mathcal{C}'}(y, y')
 =  \frac{\frac{1}{2}\,(2\pi)^{-\frac{d}{2}} \Gamma\left(\frac{d-2}{2}\right) }
{\big[\eta_{\alpha\beta}\,y^\alpha y'^\beta
+i\varepsilon\,(y^0 y'^{d+1} - y'^0 y^{d+1})\big]^{\frac{d-2}{2}}} \,,
\end{equation}
from which $D_{\sss{H}}^{+}$ can be obtained using
(\ref{eq.Wight1}). The $H=0$ case can be readily computed:
using the M-coordinates \eqref{coordMink0}-\eqref{coordMinki} we obtain 
\begin{equation}\label{dzero}
 D_{0}^{+}(x, x') =
        \frac{(-1)^{\frac{d-2}{2}}\,
\Gamma\left(\frac{d-2}{2}\right)}{2\,(2\pi)^{\frac{d}{2}}}
        \frac{1}{\big[\sigma_0 -i\varepsilon\, {\sgn}(t-t')\big]^{\frac{d-2}{2}}}
\end{equation}
with $\sigma_0 := (x - x')^2/2$ and sgn being the sign function (In fact instead of 
$\sgn(t-t')$ we get a more complex function but, at the limit
$\varepsilon \to 0^{+}$ only its sign will play a part and the latter
is found to be the same as $(t-t')$). In particular for $d=4$ one obtains the usual Wightman function:
\begin{equation}
 D_{0}^{+}(x, x') =
       \frac{-1}{8\pi^2}\frac{1}{\sigma_0 - i\varepsilon\,{\sgn}(t-t')}\,.
\end{equation}
Even more, we can obtain the de~Sitter Wightman function written in Minkowskian coordinates:
\begin{multline}\label{WdsHM}
D_{\sss{H}}^{+}(x,x')=  \left(1 - \frac{H^2}{4^{~}}x^2\right)^{(\frac{d - 2}{2})}
 \frac{(-1)^{\frac{d-2}{2}}\,
\Gamma\left(\frac{d-2}{2}\right)}{2\,(2\pi)^{\frac{d}{2}}}
\times\\
\times \frac{1}{[\sigma_0 - i\varepsilon\,{\sgn}(t-t')]^{(\frac{d - 2}{2})}}
\left(1 - \frac{H^2}{4^{~}}x'^2\right)^{(\frac{d - 2}{2})}.
\end{multline}
Note that in this formula, $\sigma_0$ is {\em not} half the squared geodesic distance in dS space.
Obviously, the formula (\ref{dzero}) is recovered directly when $H=0$.

Now let us show that the two-points function $D_{\sss{H}}^{+}$ obtained from (\ref{eq.Wight1}) also exhibits 
the Hadamard behavior in dS space. For convenience, we will carry out the calculations on $\mathcal{C}$ rather than $X_{\sss{H}}$. Each point $y$ of $X_{\sss{H}}$ is in one-to-one correspondence with the point
$y_{\sss \mathcal{C}}$ of $\mathcal{C}$ defined through
$y^\alpha = \Omega^{-1}_{\sss H} y^\alpha_{\sss \mathcal{C}}$, 
where the subscript $\mathcal{C}$ has been added to distinguish coordinates 
on the cone (\ref{coordC}) from the coordinates on $\mathcal{C'}$. Consequently,  the function 
$D^{+}_{\sss H}$ reads
\begin{equation}
D_{\sss H}^{+}(y, y')=D^{+}_{\mathcal{C}'}(y_{\sss \mathcal{C}}, y_{\sss \mathcal{C}}').
\end{equation}
 Now, using the 
definitions (\ref{PH}) and (\ref{w}) one has
\begin{equation*}
y^{d+1}_{\sss \mathcal{C}}y'^{d+1}_{\sss \mathcal{C}} - y^d_{\sss \mathcal{C}}y'^d_{\sss \mathcal{C}} 
= -\frac{w_{\sss \mathcal{C}}}{H}\,\frac{w'_{\sss \mathcal{C}}}{H} + H^{-2}.
\end{equation*}  
Note that since the constraint coming from (\ref{PH}) has been taken into account the coordinates
$\{y^\mu_{\sss \mathcal{C}}, w_{\sss \mathcal{C}} H^{-1} \}$ are nothing but the usual ambient space coordinates for the dS space.
Introducing the function (see for instance \cite{allen, allenjacobson})
\begin{equation*}
\mathcal{Z} = \cosh\bigl(\mu(y,y') H\bigr) , 
\end{equation*}
where $\mu(y,y')$ is the geodesic distance on $X_{\sss H}$, and reminding the relation
\begin{equation*}
y^0_{\sss \mathcal{C}}y'^0_{\sss \mathcal{C}} - 
\boldsymbol{y}_{\sss \mathcal{C}}\cdot\boldsymbol{y'}_{\sss \mathcal{C}} - \frac{w_{\sss \mathcal{C}}}{H}\frac{w'_{\sss \mathcal{C}}}{H} = - H^{-2} \mathcal{Z},
\end{equation*}  
one obtains
\begin{equation*}
\eta_{\alpha \beta} y^\alpha_{\sss \mathcal{C}} y'^\beta_{\sss \mathcal{C}} = - H^{-2} (\mathcal{Z} - 1).
\end{equation*}  
Finally, the two-points Wightman function in dS space reads
\begin{equation}
 D^{+}_{\sss H}(y, y')
 =  \frac{(-1)^{\frac{d-2}{2}}\frac{1}{2}\,(2\pi)^{-\frac{d}{2}}\Gamma\left(\frac{d-2}{2}\right)}
{\big[H^{-2} (\mathcal{Z} - 1)
- i\varepsilon\,(y^0_{\mathcal{C}} y'^{d+1}_{\mathcal{C}} - y'^0_{\mathcal{C}} y^{d+1}_{\mathcal{C}})\big]^{\frac{d-2}{2}}} \,,
\end{equation}
The relation
\begin{equation*}
\lim_{\mu \to 0}  H^{-2} (\mathcal{Z} - 1) =
\sigma_{\sss H} , 
\end{equation*} 
where $ 2 \sigma_{\sss H} = \mu^2$, ensures that $D^{+}_{\sss H}$ has the Hadamard behavior. As a consequence, the vacuum of our theory is the Euclidian one.

\section{The $\Large d=2$ case}\label{d=2}
Most of the previous results are still valid for the $d=2$ case. In particular, the correspondence (\ref{phi-phih}) now reads
\begin{equation}
\phi^\xi=\phi.
\end{equation}
Nevertheless this case is so much peculiar that it requires a special treatment. The principal new feature 
is that the space of solution of (\ref{eqpc}) is not an Hilbert space (due to the infrared divergence problem).
The other peculiarity comes from the splitting of $so(2,2)$ into two independent parts. 

The manifold $X_{\xi}$ is defined without change. We now consider the projective cone $S^1\times S^1/\setZ_{\sss 2}$ rather than ${\cal C}'$, this is coherent with the end of the section IV. 
This projective cone can be written as the following points of $\setR^4$:
\begin{equation}\label{coordC2}
\left \{
 \begin{array}{lcl}
  y^{3} &=& \cos \beta \\
  y^0 &=&  \sin \beta \\
  y^1 &=& \sin \alpha\\
  y^2 &=& \cos \alpha,
 \end{array}
\right .
\end{equation}
in which we have identified $(\alpha,\beta)$ with $(\alpha+\pi,\beta+\pi)$.
This allows to set
\begin{equation}
\left\{
\begin{array}{rcl}
u^+&=&\alpha+\beta\\
u^-&=&\alpha-\beta
\end{array}\right.
\end{equation}
which realizes the isomorphism
\begin{equation}
\begin{array}{rcl}
S^1\times S^1/\setZ_{\sss 2}&\to&S^1\times S^1\\
(\alpha,\beta)&\mapsto&(u^+,u^-).
\end{array}
\end{equation}
\smallskip
In these new coordinates the splitting $so(2,2)=so(1,2)\oplus so(1,2)$ becomes very clear because of the following realizations:
\begin{align*}
\frac{1}{2}(X_{12}+\varepsilon X_{03})&=\frac{\partial}{\partial u^{\varepsilon}}\\
\frac{1}{2}(X_{32}+\varepsilon X_{01})&=-\sin u^{\varepsilon}\frac{\partial}{\partial u^{\varepsilon}}\\
\frac{1}{2}(X_{13}+\varepsilon X_{02})&=-\cos u^{\varepsilon}\frac{\partial}{\partial u^{\varepsilon}},
\end{align*}
in which $\varepsilon=\pm1$. One can verify readily that the above combinations
are two copies, one for each value of $\varepsilon$, of the lie algebra $so(1,2)$.
The equation (\ref{eqpc}) reads
\begin{equation}
\frac{\partial^2 \phi}{\partial u^+\partial u^-}=0,
\end{equation}
whose general solution reads $\phi(u^+,u^-)=f(u^+)+g(u^-)$.
The scalar product is, in this case, degenerate and the space of solutions splits into two spaces carrying an indecomposable representation of the group SO(1,2).  Details can be found in \cite{dbr3, dbr2, dbr4}.

\section*{Acknowledgments}
The authors thank  J-P. Gazeau for valuables 
discussions and comments.

%------

\begin{thebibliography}{AAA} \baselineskip=10pt
\bibitem{birreldavis} N.D.~Birrell and P.C.W.~Davies, \emph{Quantum fields in
curved space}. Cambridge University Press, 1982.
\bibitem{fulling} S.A.~Fulling \emph{Aspects of quantum field theory in curved spacetime}.
Cambridge University Press, 1989.
\bibitem{wald} R.M.~Wald, \emph{Quantum field theory in curved spacetime and black hole 
thermodynamic}. The University of Chicago Press, Chicago, 1994.
\bibitem{bk} J. Bi\v{c}\'ak and P. Krtou\v{s}, Phys. Rev. {\bf D64}  124020 (2001).
\bibitem{barutraczka} A.O.~Barut and R.~Raczka , \emph{Theory of group representations and applications}.
World Scientific Publishing, 1986.
\bibitem{fronsdal} C. Fronsdal, Phys. Rev. {\bf D12}, 3819 (1975).
\bibitem{onofri} E.~Onofri, J. Math. Phys. {\bf 17}, 401 (1976). 
\bibitem{zg} B. Zhou and H-Y. Guo,  hep-th/0512235.
\bibitem{jpg} J-P. Gazeau, {\it Repr\'esentations des groupes et semi-groupes dans le probl\`eme de l'atome d'hydrog\`ene}, unpublished.
\bibitem{ghr1} T. Garidi, E. Huguet, and J. Renaud, Phys. Rev.   {\bf D67}, 124028 (2003).
\bibitem{Hyp.Geom.85} Z.-Y.~Wen and J. Avery,  J. Math. Phys. {\bf 26} 3, (1985).
%{\em Some properties of hyperspherical harmonics},
\bibitem{allen} B. Allen, Phys. Rev. {\bf D32}, 3136 (1985).
\bibitem{allenjacobson} B. Allen, T. Jacobson,  Comm. Math. Phys. {\bf 103}, 669 (1986).
\bibitem{dbr3} S. De Bievre and J. Renaud, Phys. Rev. {\bf D57}, 6230 (1998).
\bibitem{dbr2} S. De Bievre and J. Renaud, Lett. Math. Phys. {\bf 34}, 385 (1995).
\bibitem{dbr4} S. De Bievre and J. Renaud, J. Phys. {\bf A34}, 10901 (2001).
\end{thebibliography}
\end{document}